\documentclass[twocolumn,showpacs, amsmath,amssymb]{revtex4}
\usepackage{amssymb}
\usepackage[dvips]{graphicx}
\usepackage{color}
\begin{document}

\title{Magnetoresistance in organic spintronic devices: the role of nonlinear effects.}
\author{A. V. Shumilin$^1$, V.V. Kabanov$^2$, and V.A. Dediu$^3$}

\affiliation{$^1$A.F.Ioffe Physico-Technical Institute, St.-Petersburg 19
4021, Russia.\\
$^2$Department for Complex Matter, Jozef Stefan Institute, 1001 Ljubljana, Slovenia.\\
$^3$CNR - ISMN, Consiglio Nazionale delle Ricerche - Istituto per lo Studio dei Materiali Nanostrutturati, v. Gobetti 101, 40129, Bologna, Italy}

\begin{abstract}
We derive kinetic equations describing injection and transport of spin polarized carriers in organic semiconductors with hopping conductivity via an impurity level. The model predicts a strongly voltage dependent magnetoresistance, defined as resistance variation between devices with parallel and antiparallel electrode magnetizations (spin valve effect). The voltage dependence of the magnetoresistance splits into three distinct regimes. The first regime matches well known inorganic spintronic regimes, corresponding to barrier controlled spin injection or the well known conductivity mismatch case. The second regime at intermediate voltages corresponds to strongly suppressed magnetoresistance. The third regime develops at higher voltages and accounts for a novel paradigm. It is promoted by the strong non-linearity in the charge transport which strength is characterized by the dimensionless parameter $eU/k_BT$. This nonlinearity, depending on device conditions, can lead to both significant enhancement or to exponential suppression of the spin valve effect in organic devices.
We believe that these predictions are valid beyond the case of organic semiconductors and should be considered for any material characterized by strongly non-linear charge transport.

\end{abstract}


\maketitle

\section{Introduction}

Magnetoresistance in organic spintronic devices is an easy effect to detect \cite{Dediu2009}.
 Vertical multilayered  devices combining manganite (La$_{0.7}$Sr$_{0.3}$MnO$_3$) and cobalt external magnetic
electrodes with an Alq$_3$ transport interlayer have perhaps never failed (no negative reports
available) to exhibit a clearly measurable magnetoresistance
\cite{Xiong2004,Xu2007,Vinzelberg2008,Dediu2008,Sun2010}. This magnetoresistance, often called in
literature giant magnetoresistance (GMR) or spin valve effect (SV), is generally measured as a
difference between the device resistances for antiparallel and parallel electrode
magnetizations, where the latter is expected to facilitate the transport of spin polarized
currents. Manganite based devices commonly show an inverse spin valve effect featuring lower
resistance for antiparallel orientation. This property was explained
 recently by the spin filtering effects at the organic-ferromagnetic interface
\cite{Barraud2010,Steil2013,Dediu2013}.
Substituting the manganite with a 3d metal or other magnetic electrode leads to a more
 controversial picture merging successes and failures. Generally, this second case has well
 established achievements in the tunneling regime, that is for a few nm thick layer of various
 organic semiconductors \cite{Moodera,Coey}. On the other hand, the application of two metallic
  magnetic electrodes in combination to thick organic layers, expected to transport carriers via
 molecular electronic states, has lead to a number of reports claiming no
 magnetoresistance(see
  for example \cite{Jiang2008} and high thicknesses regimes in \cite{Moodera,Coey})
  and a number of positive communications \cite{Wang2005,Seneor_Mattana_unpub}.

An important fact to underline is the voltage dependence of the GMR in organic
based devices.
The magnetoresistance is generally detectable at relatively small voltages ($\le 0.1$ V)
which are much lower than interfacial electronic
barriers for the injection into two transport states of OSC, HOMO and LUMO.  Indeed, typical values of
the barriers measured by various spectroscopic techniques are of the order of
1 eV \cite{Dediu2008}.
This clearly implies the need to consider eventual intragap states or band, as it is discussed in \cite{dediu_riminucci2013,Yu2014}. It is also known from experiments that impurity states
can be created in OSC in uncontrollable way during sample preparation \cite{rybicki2012}.

Such easily achievable spintronic effects in electrically operated organic devices looked strange
especially when compared to inorganic counterpart, where the fight with materials and concepts
impeded similar achievements for years of tough and extremely qualified research.
For inorganic
devices these difficulties were convincingly explained through the conductivity mismatch problem
\cite{schmidt,van_Son,Fert_Jaffres}. A fundamental obstacle for the spin injection from a ferromagnetic metal to a
semiconductor
was formulated by Schmidt et. al.\cite{schmidt}. It was shown that in the
case of diffusive transport the large difference in
conductivities of ferromagnetic metal and semiconductor limits the spin
injection coefficient $\gamma \propto \sigma_s/\sigma_F$,
where $\sigma_s,\sigma_F$ are conductivities of semiconductor and ferromagnetic
metal respectively. This limitation can be overcome by introducing a tunneling barrier at
the ferromagnetic-semiconductor interface \cite{rashba,Fert_Jaffres}, leading to an essential improvement of the spin
injection into a semiconductor. In order to be effective the tunnel resistance
should be of the order of the effective resistance defined as
$L_F/\sigma_F,L_s/\sigma_s$, where $L_F,L_s$ are the spin diffusion lengths in ferromagnetic metal and semiconductor respectively.

The question whether OSC are free of conductivity mismatch or the MR in OSC may not be caused by real spin injection into the organic electronic states has raised. The latter doubt was strongly enhanced by the absence of Hanle effect in all the tested organic spintronic devices\cite{Riminucci,schmidt_hanle}. Indeed in a device based on the transport of spin polarized carriers, the application of a magnetic field perpendicular to the carrier's spin causes the latter to precess with a frequency 28 GHz per Tesla, regardless of whether the transporting medium is inorganic or organic. This precession causes a misalignment between the relative orientation of the injected spin polarization and the electrodes' magnetization, resulting in a measurable resistance change. Noteworthy, in inorganic spintronics Hanle effect has acquired the role of decisive proof for the demonstration of a spin polarized electrical injection in the investigated medium \cite{Johnson_Silsbee,Monzon}.

While interesting and stimulating ideas have already been advanced to explain the absence of the Hanle observation \cite{Yu2013}, we would like to underline an important paradox related to organic spintronic devices. Although magnetoresistance represents the parameter whose variation is used to indicate the strength of the Hanle effect, there exist not yet a clear understanding of the nature of magnetoresistance in such devices.

At present there were few attempts to model spin transport in organic semiconductors\cite{deJong,smith,Barraud2010}. The first successful approaches to describe magneto-resistance in OSC were based on the extension of the models describing inorganic devices \cite{rashba,Fert_Jaffres} through an accurate inclusion of the mechanisms of charge injection in OSC\cite{smith}. An important step forward was proposed in Refs.\cite{deJong,koopmans}. The authors have considered a model of multistep tunneling through single molecular level in OSC and the finite occupation of the level during tunneling process. It promoted thus for the first time elements of nonlinearity for the spin polarized carrier transport inside OSC. Nevertheless, non linear effects were considered only for a small number of intermediate states, while for the ÒbulkÓ hopping case the resistance of organic was described as a constant Refs.\cite{deJong}.

Here we make a step further and consider charge transport nonlineartiy in an organic semiconductor confined between two spin polarized electrodes. We limit our model to the conductivity via a single intragap electronic level. In spite of its obvious simplification (compared to strongly non-homogeneous situation in OSC) the model clearly confirms two well known spintronic regimes described above, and namely the conductivity mismatch \cite{schmidt} and the tunnel barrier regime \cite{rashba,Fert_Jaffres}, as expected from both experimental results and theoretical predictions. Surprisingly we discover a third regime, explicitly characteristic for organic semiconductors, where non linear effects induce a measurable magneto-resistance. Indeed, for strongly voltage dependent hopping transport through the molecular level, the conductivity becomes strongly dependent on the injected spin. We demonstrate
that in high voltage limit $eU/k_BT>>1 $ the resistance of the organic semiconductor is governed by the contacts. Here $e$ is the elementary charge, $U$ is the voltage, $k_B$ is the Boltzmann constant, and $T$ is the temperature. In that case boundary conditions
define the resistivity of organic semiconductor and therefore the resistivity becomes spin dependent. It means that contrary to the case of ordinary semiconductor with diffusive transport the conductivity mismatch arguments are applicable for organic semiconductors
only for some limited voltage range, while for higher voltages strong nonlinear effects in organic semiconductor enable the detection of the magnetoresistance caused by spin polarized injection.

Although we have formulated the model for the conductivity via the impurity level, it is valid for
any energy level with hopping conductivity when the space-charge may be neglected. As a result the
model is valid if the conductivity occurs over the HOMO and LUMO provided that concentration of the injected carriers is small. Experimental situation as far as
transport in SV devices is concerned remains controversial. The transport due to electron and hole injection into the HOMO and LUMO is commonly assumed in literature\cite{smith,bobbert}. On the other hand the I-V characteristics of the device in that case should be highly nonlinear and sensitive to the temperature\cite{Yu2014}. This was not reported experimentally.
These arguments lead to the suggestion that the conductivity of the SV devices is due to existence  of some impurity levels when the applied voltages are much less than Fermi level-LUMO (and Fremi level - HOMO) distance \cite{dediu_riminucci2013}. The origin of the impurity levels may be different (for example metal atoms, oxygen molecules \cite{Yu2014} or X-rays induced
traps during electron beam deposition of metallic electrodes \cite{rybicki2012}) . Contrary to Ref.\cite{Yu2014} where it was suggested that impurity levels have wide energy distribution between the HOMO and LUMO we assume that the impurity level has narrow energy distribution.
We show in our work that for such
levels the non-linearity can both suppress but also enhance spin valve effect in some cases.

\section{Main Equations}

In order to describe nonlinear effects we assume that the transport in an organic semiconductor in the low voltage limit
($eU << \Delta$, $\Delta$ is the LUMO-HOMO splitting)
is determined by the polaron hopping over impurity levels similar to Ref.\cite{Yu2013}. In order to have reasonable conductivity the level should have energy close to the Fermi energy.
We assume that the density of this levels $n$ is relatively small ($\leq 10^{18} cm^{-3}$)and therefore
we can neglect the electric field which appears due to finite occupancy of the impurity levels by polarons.
The electric field inside the semiconductor is equal to the external field in that case.
We also assume that spatial energy fluctuations of these impurity levels are small $\Delta \epsilon << k_BT$.

To describe the polaron hopping transport we apply the stationary kinetic equation for the distribution function
$f_{\sigma}(m)$ in the site representation derived in Ref.\cite{bryksin}:
\begin{eqnarray}
0&=&\sum_{m'}\Bigl(f_{\sigma}(m')(1-f_{\sigma}(m)))W_{m',m} \cr
&-&f_{\sigma}(m)(1-f_{\sigma}(m'))W_{m,m'}\Bigr)
\label{kinetik1}
\end{eqnarray}
where
\begin{equation}
W_{m',m}=\exp{(\beta eE(m'-m))}w(|m'-m|)
\end{equation}
and $w(|n|)={J(|n|)^2\sqrt{\pi\beta}\exp{(-\beta E_a)}\over{2\hbar\sqrt{E_a}}}$. Here $\beta=1/k_BT$, $E_a$ is activation energy of polaron,
$\hbar$ is the Planck constant, $e$ is the charge of electron, and $J(|n|)$ overlap integral between two sites. This equation are derived under condition that the spin relaxation is absent
and the Hubbard energy is neglected.
We consider only the case
of weak field $\beta eEa <<1$ ($a$ is the hopping distance), therefore we can keep only lowest order in the expansion of the hopping probability in electric field $E$:
$W_{m',m}=w(|m'-m|)(1+\beta eE(m'-m))$.  Assuming that distribution function $f_{\sigma}(m)$ is smooth function of $m$, we may expand
$f_{\sigma}(m')=f_{\sigma}(m)+(m'-m){d\over{dm}}f_{\sigma}(m)+{(m'-m)^2\over{2}}{d^2\over{dm^2}}f_{\sigma}(m)$ and consider $m$ as a
continuous variable $x$. In that case Eq.(\ref{kinetik1}) reads:
\begin{equation}
D{d^2f_{\sigma}(x)\over{dx^2}}+\beta e E D{d\over{dx}}\bigl(f_{\sigma}(x)(1-f_{\sigma}(x))\bigr)=0
\label{kinetik2}
\end{equation}
Here the spin independent diffusion coefficient is defined as $D=\sum_{n}w(n)n^2a^2/6$. $\beta eD f_{\sigma}(1-f_{\sigma})$ is the local conductivity of polarons with the spin $\sigma$.
The first integrals of these equations ( \ref{kinetik2}) define the currents of spin "up" $J_{\uparrow}$ and spin "down" $J_{\downarrow}$ polarons:
\begin{eqnarray}
{J_{\uparrow}\over{en}}&=&D{df_{\uparrow}(x)\over{dx}}+\beta e E D f_{\uparrow}(x)(1-f_{\uparrow}(x))\cr
{J_{\downarrow}\over{en}}&=&D{df_{\downarrow}(x)\over{dx}}+\beta e E D f_{\downarrow}(x)(1-f_{\downarrow}(x))
\label{kinetik3}
\end{eqnarray}
Here $n$ is density of conducting levels.
These equations are similar to ordinary diffusion equation describing spin-polarized
currents in ordinary semiconductors\cite{YuFlatte,Pershin}. The main difference is in the dependence of the drift current on the distribution function.
If the distribution function is small the current is defined by $f <<1$. If the level is filled by electrons
the current is determined by the concentration of holes $(1-f)$. The later term is missing in the ordinary description of the spin
transport in semiconductors described by ordinary diffusion equation.
These equations do not include the spin relaxation term (its effect will be discussed below). This holds if the thickness of the organic layer $L < \sqrt{D\tau_s}$ where $\tau_s$ is the spin relaxation time in organic semiconductor. In the case
of organic semiconductor $\tau_s\simeq 10^{-3}-10^{-6}s$\cite{sanvito} is large
and $\sqrt{D\tau_s}$ is of the order of few hundred nanometers.

Eqs.(\ref{kinetik3}) should be supplied by the boundary conditions, which describe the microscopic
structure of the contacts. Here we consider an oversimplified model (similar to that proposed in
Ref.\cite{rashba})that assumes the absence of the spin relaxation in the contacts.  The change of
electro-chemical potential $\delta \xi$ in the contact is approximate by the Ohm's law
$\delta \xi_{l,r;\sigma} = J_\sigma R_{l,r;\sigma}$. Here  index $r,l$ stands for right and left
interface and $\sigma =\uparrow,\downarrow$ stands for "up" and "down" spin polarization. $R_{l,r,\sigma}$ are
the contact resistivities for up and down spin.

Differently from Ref.\cite{rashba}, in our case it is important to consider
not only the change of the
electro-chemical potential $\Delta \xi_{l,r;\uparrow,\downarrow}$, but also the contribution to it from the
contact voltage. Note that the organic side of the
contact cannot have surface charge, therefore the electric field in the contact is the same as in
 the bulk of organic layer. It means that we can neglect the contact voltage provided that
contact is much shorter then the thickness of the organic layer. Therefore the change of the
electrochemical potential $\delta \xi_{l,r;\sigma}$ is equal to the change of chemical potential.
It means that the boundary conditions for the
distribution function at the interface are defined as follows:

\begin{eqnarray}
&&f_{\uparrow,\downarrow}(0)={1\over{\exp{(\beta(\mathcal{E}_0-eR_{l,\uparrow,\downarrow}J_{\uparrow,\downarrow}))}+1}}, \cr
&&f_{\uparrow,\downarrow}(L)={1\over{\exp{(\beta(\mathcal{E}_0+
eR_{r,\uparrow,\downarrow}J_{\uparrow,\downarrow}))}+1}}
\label{boundary}
\end{eqnarray}
Here $\mathcal{E}_0$ is the position of the energy level in organic semiconductor with respect to the Fermi level.
As a results Eq.(\ref{kinetik3}) define distribution function $f_{\uparrow,\downarrow}(x)$
inside the organic layer. The substitution of this solution to the boundary conditions Eq.(\ref{boundary})
allows us to obtain the equation for the current  $J_{\uparrow,\downarrow}$

In order to discuss the current-voltage characteristics and the spin-valve effect we rewrite equations (\ref{kinetik3}) in dimensionless form:
\begin{equation}
{df_{\uparrow,\downarrow}(y)\over{dy}}+f_{\uparrow,\downarrow}(y)
(1-f_{\uparrow,\downarrow}(y))=j_{\uparrow,\downarrow}/\zeta
\label{kindl}
\end{equation}
where dimensionless current defined as $j_{\uparrow,\downarrow}=J_{\uparrow,\downarrow}L/enD$, and dimensionless voltage
$\zeta=\beta eEL$, $y=\alpha x$, $\alpha=\zeta/L$
General solution of Eq.(\ref{kindl}) is:
\begin{equation}
f_{\uparrow,\downarrow}(y)={1\over{2}}-{f_{1\uparrow,\downarrow}-
f_{2\uparrow,\downarrow}\over{2}}\tanh{(\kappa(y-y_{0\uparrow,\downarrow}))}
\label{soldl}
\end{equation}
Here $f_{1\uparrow,\downarrow}={1\over{2}}(1+\kappa)$, $f_{2\uparrow,\downarrow}={1\over{2}}(1-\kappa$) and $\kappa=(1-4j_{\uparrow,\downarrow}/\zeta)^{1/2}$.
At the fixed voltage $\zeta$ currents $j_{\uparrow,\downarrow}$ and parameter $y_{0\uparrow,\downarrow}$ are defined from the boundary conditions:
\begin{eqnarray}
&&f_{\uparrow,\downarrow}(0)=f_{L\uparrow,\downarrow}=
{1\over{\exp{(\epsilon_0-r_{l\uparrow,\downarrow}j_{\uparrow,\downarrow}))}+1}}, \cr
&&f_{\uparrow,\downarrow}(\zeta)=f_{R\uparrow,\downarrow}=
{1\over{\exp{(\epsilon_0+r_{r\uparrow,\downarrow}j_{\uparrow,\downarrow}))}+1}}
\label{bounddl}
\end{eqnarray}
where we introduce the dimensionless position of the energy level in organic semiconductor $\beta\mathcal{E}_0=\epsilon_0$  and the dimensionless
contact resistances defined as $r_{r,l\uparrow,\downarrow}=R_{r,l\uparrow,\downarrow}/R_o$ and the resistance of organic layer is $R_o=L/\beta e^2 n D$. Unfortunately
the accurate solutions of these equations is possible only by numerical techniques. Some accurate analytical solutions are
possible only in some limiting cases. The spin valve magneto-resistance is defined as follows:
\begin{equation}
MR={j_P-j_A\over{j_A+j_P}}
\label{spinwal}
\end{equation}
where $j_{A,P}$ is the current of the device with parallel and antiparallel polarization of the ferromagnetic electrodes with the fixed voltage.
For practical purposes it is convenient to exclude $\exp{(-y_{0\uparrow,\downarrow})}$ from Eq.(\ref{bounddl}). As result we obtain a single transcendental
equation for the current $j$:
%\begin{eqnarray}
%&&f_{L\uparrow,\downarrow}(f_{1\uparrow,\downarrow}-f_{2\uparrow,\downarrow}\exp{(\tilde{\zeta})})-f_{1\uparrow,\downarrow}f_{2\uparrow,\downarrow}(1-\exp{(\tilde{\zeta})})=\cr
%&&f_{L\uparrow,\downarrow}f_{R\uparrow,\downarrow}(1-\exp{(\tilde{\zeta})})-f_{R\uparrow,\downarrow}(f_{2\uparrow,\downarrow}-f_{1\uparrow,\downarrow}\exp{(\tilde{\zeta})})
%\label{cvceq}
%\end{eqnarray}
\begin{eqnarray}
&&(f_{1\uparrow,\downarrow} - f_{L\uparrow,\downarrow})(f_{R\uparrow,\downarrow} - f_{2\uparrow,\downarrow}) = \cr
&&e^{-\tilde{\zeta}} (f_{1\uparrow,\downarrow} - f_{R\uparrow,\downarrow})(f_{L\uparrow,\downarrow} - f_{2\uparrow,\downarrow}).
\label{cvceq}
\end{eqnarray}
Here $\tilde{\zeta}=\kappa\zeta$.

\begin{figure}
\includegraphics[width = 87mm, angle=-0]{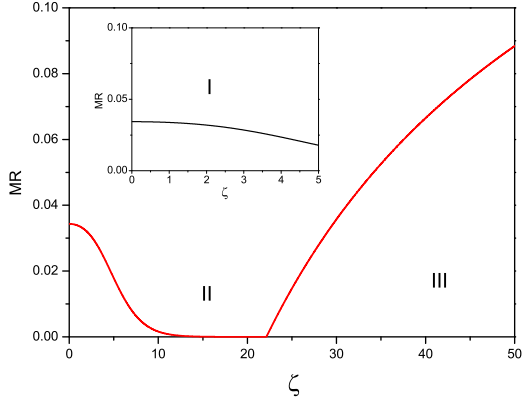}
\caption{Spin-valve magneto-resistance $MR$ as a function of dimensionless voltage $\zeta$. $\epsilon_0=2$,
$r_\uparrow=1$ $r_\downarrow=5$. Organic layer resistance $r_o=9.52$. The low voltage region (I) corresponds to the standard theory
Refs.\cite{schmidt,Fert_Jaffres,rashba}. In the intermediate voltage region (II) the spin-valve is exponentially reduced. Region III corresponds
to recovered spin-valve effect in the strong voltage region.}
\end{figure}

\section{Magnetoresistance}

The results of numerical solution of Eqs.(\ref{cvceq}) for currents and spin valve magneto-resistance  Eq.(\ref{spinwal}) are plotted in Fig.1.
As it follows from this figure the spin-valve magneto-resistance has three characteristic regions. At small voltages $eU<<k_BT$ we expand in Eq.(\ref{cvceq})
$f_{L,R\uparrow,\downarrow}$ up to the first order in $r_{r,l\uparrow,\downarrow}j_{\uparrow,\downarrow} <<1$. After straightforward calculations we find:
\begin{equation}
j_{\uparrow,\downarrow}={\zeta\over{r_{l\uparrow,\downarrow}+r_{r\uparrow,\downarrow}+r_o}}
\label{lowE}
\end{equation}
where the resistance of organic layer is defined as $r_o=4\cosh^2{(\epsilon_0/2)}$.
Therefore in the limit of low field spin-valve magneto-resistance $MR$ is governed by the conductivity
mismatch condition and by the contact resistances. Indeed in this limit we are able to
reproduce well known results for ordinary semiconductors\cite{schmidt,Fert_Jaffres,rashba}.
In the symmetric case
$r_{l\uparrow,\downarrow}=r_{r\uparrow,\downarrow}=r_{\uparrow,\downarrow}$  spin-valve magneto-resistance is given by the formula:
\begin{equation}\label{MR0U}
MR={(r_\uparrow-r_\downarrow)^2\over{(r_\uparrow+r_\downarrow+r_o)^2+
(2r_\uparrow+r_o)(2r_\downarrow+r_o)}}.
\end{equation}
Therefore in the case when total resistance is governed by the resistance of the organic layer
we have $MR={(r_\uparrow-r_\downarrow)^2\over{2r_o^2}} \ll1$ in agreement with Refs.\cite{schmidt,Fert_Jaffres}. For large enough contact resistances Eq.(\ref{MR0U}) gives a measurable magnetoresistance in agreement with the results of Refs.\cite{Fert_Jaffres,rashba}.

The spin-valve magneto-resistance decreases with applied voltage, because effective resistivity increases with $U$.
In the limit of intermediate voltage $eU \gg k_B T$ ($\tilde{\zeta} \gg 1$) the nonlinear effects become important. The right-hand side of Eq. (\ref{cvceq})
is exponentially small $\exp(-\tilde{\zeta})\ll 1$ and may be neglected. In that case Eq. (\ref{cvceq})
has two roots leading to the transcendental equations for the current:
$$
f_{1\uparrow,\downarrow}=f_{L\uparrow,\downarrow} \, \Rightarrow \, j_{L\uparrow,\downarrow}={\zeta\over{\exp{(-\epsilon_0+r_{l\uparrow,\downarrow}j_{L\uparrow,\downarrow})}+1}}
\label{strongE}
$$
$$
f_{2\uparrow,\downarrow}=f_{R\uparrow,\downarrow} \, \Rightarrow \, j_{R\uparrow,\downarrow}={\zeta\over{\exp{(\epsilon_0+r_{r\uparrow,\downarrow}j_{R\uparrow,\downarrow})}+1}}
$$
They have two solutions $j_{L\uparrow,\downarrow}$ and $j_{R\uparrow,\downarrow}$ respectively.
The correct solution should satisfy the condition $f_{1\uparrow,\downarrow} \ge f_{L\uparrow,\downarrow}> f_{R\uparrow,\downarrow}\ge f_{1\uparrow,\downarrow}$. These inequalities are satisfied if
$j_{\uparrow,\downarrow}=min(j_{R\uparrow,\downarrow},j_{L\uparrow,\downarrow})$.  This formula defines current voltage characteristic in the large current limit.

For the intermediate field case, when $k_B T \ll eU \ll \mathcal{E}_0$ ($1\ll\tilde{\zeta}\ll\epsilon_0$), the correct root is determined by the sign of $\epsilon_0$. In this case the conductivity is determined by the contact with positve energetic mismatch and the device shows no spin-valve effect. The actual value of the spin-valve magneto-resistance $MR$ is exponentially small function of the applied voltage $\exp(-\tilde{\zeta})$.

However at higher voltages when $jr_{lr,\uparrow,\downarrow} \gg \epsilon_0$ the correct root is determined not by the position of the organic level $\epsilon_0$, but by the largest of the contact resistances. In the antiparallel (AP) configuration the root is always defined by the resistance $r_\uparrow$. Here we again assume that $r_\uparrow>r_\downarrow$. On the other hand in the parallel (P) configuration for one of spin directions the boundary with the resistance $r_\uparrow$ is absent and the root is determined by $r_\downarrow$. It leads to reappearance of the spin-valve magneto-resistance at large voltage.
In order to clarify these processes we plot in Fig.2 and Fig. 3 the spatial dependence of the level occupancy and the chemical potential in the organic level for the parallel and antiparallel configurations. As it is clearly seen, the resistivity of the device is determined by small density of holes in the left part of the organic layer and by the small density of electrons in the right part of the layer. Therefore the resistances of the antiparallel configuration is equal to the resistance of the parallel configuration $P\uparrow$. The parallel configuration $P\downarrow$ has lower resistance leading to a large magnetoresistance.
\begin{figure}
\includegraphics[width = 87mm, angle=-0]{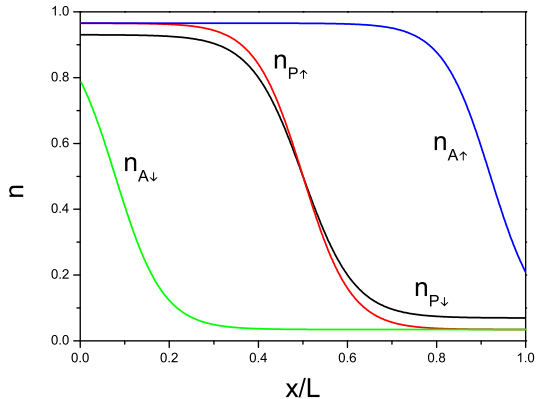}
\caption{Spatial dependence of the level occupancy for spin up $\uparrow$ and spin down
$\downarrow$ electrons in the device with parallel "P" and antiparallel "A" configurations of contacts.}
\end{figure}

\begin{figure}
\includegraphics[width = 87mm, angle=-0]{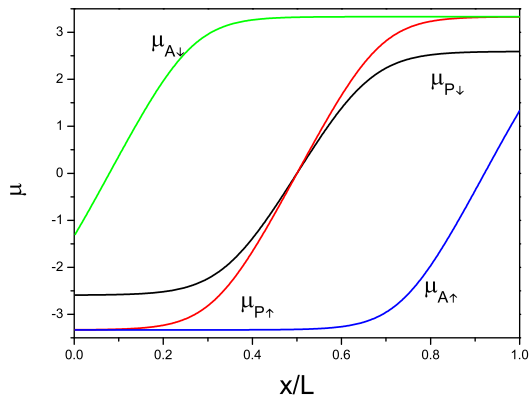}
\caption{Spatial dependence of the chemical potential for spin up $\uparrow$ and spin down
$\downarrow$ in the device with parallel "P" and antiparallel "A" configurations of contacts.}
\end{figure}

With the further increase of the voltage the equation for current may be solved with
the logarithmic accuracy $j_{L,R\uparrow,\downarrow}=r^{-1}_{l,r\uparrow,\downarrow}\ln{(\zeta)}$. In this limit the spin-valve magneto-resistance
increases, logarithmically approaching the limiting value:
\begin{equation}\label{limit}
MR={(r_\uparrow-r_\downarrow)\over{(r_\uparrow+3r_\downarrow)}}
\end{equation}
which is larger then weak field value (\ref{MR0U})  This formula does not contain any information about the resistance of the
organic layer and therefore represents the perfect device.

We have described above three limiting cases for the spin valve effect. It is important to underline that the position of the conducting level  $\epsilon_0$ with respect to
the chemical potential of ferromagnetic contacts is a key parameter for establishing a magnetoresistance (spin valve effect). Figure 4 shows the dependence of the magnetoresistance on the difference between level energy and chemical potential of the ferromagnetic layer $\epsilon_0$ for constant $r_\uparrow,_\downarrow$ resistances.
When $\epsilon_0<1$  the spin-valve effect monotonically grows with applied voltage reaching
the asymptotic value Eq.(\ref{limit}) (Fig.4). When $\epsilon_0= 1.5$ the minimum of spin-valve
effect at intermediate valuees of electric fields $\zeta$. At $\epsilon_0 = 2$ we clearly see the three regions, including that where the spin valve effect is fully suppressed.

\begin{figure}
\includegraphics[width = 87mm, angle=-0]{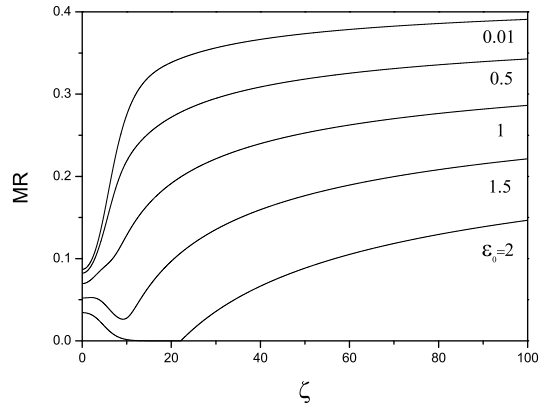}
\caption{Spin-valve magneto-resistance $MR$ as a function of dimensionless voltage $\zeta$ for $r_\uparrow=1$ $r_\downarrow=5$ and different positions of conduction level $\epsilon_0$.}
\end{figure}

\section{Spin relaxation}

Neglecting the spin relaxation (as it is done in Ref.\cite{rashba}) it is possible to suppress conductivity missmatch only by introducing the contact resistances. When the contact resistances become larger than the resistance of the normal layer, the device is not subject of the conductivity mismatch and high SV effect can be achieved.

This is not the case if even a weak spin relaxation is included. It was shown in \cite{Fert_Jaffres, ValetFert} that large contact resistances lead to strong spin accumulation and depolarization near the boundary. So the contact resistances should be selected in a narrow range to avoid both phenomena: spin depolarization and conductivity missmatch.

Indeed our model differs from one discussed in \cite{rashba} and \cite{Fert_Jaffres}. In our case the conductivity missmatch is suppressed not only by the contact resistance but also due to applied voltage and non-linearity. Also we have some limit for spin accumulation because in our case the number of electronic states is finite and we can not have $f_{\uparrow}$ or $f_{\downarrow}$ larger than unity.

Let us now include in our equations a spin relaxation time $\tau$ that does not depend on applied voltage
$$
D{d^2f_{\sigma}(x)\over{dx^2}}+\beta e E D{d\over{dx}}\bigl(f_{\sigma}(x)(1-f_{\sigma}(x))\bigr)= \frac{f_\sigma(x) - f_{-\sigma(x)}}{\tau}.
$$

After introducing spin currents in the dimensionless form we obtain
$$
\frac{d j_\uparrow}{dy} = - \frac{d j_\downarrow}{dy} = \frac{\gamma}{\zeta} (f_\uparrow - f_\downarrow), \quad \gamma = \frac{L^2}{D\tau};
$$
$$
\frac{d f_\uparrow}{dy} + f_\uparrow (1-f_\uparrow) = j_\uparrow/\zeta
$$
$$
\frac{d f_\downarrow}{dy} + f_\downarrow (1-f_\downarrow) = j_\downarrow/\zeta
$$
This set of equations is much more complex then equations without spin relaxation and allows only
numerical analysis. Let us also note that the perturbation theory for small $\gamma$
 is not useful in our case because the correct perturbation parameter appears to be
 $\gamma e^\zeta$ and quickly becomes large at hight voltages. Therefore we solve these
 equations numerically for the case $\epsilon_0 = 0$.

\begin{figure}[!hbp]
\includegraphics[width = 87mm]{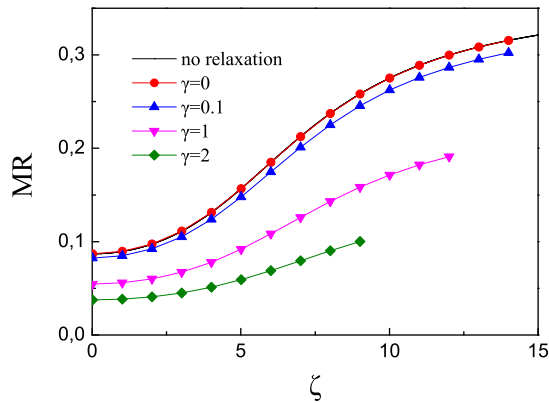}
\caption{The magnetoresistance for the device with different spin relaxation rates $\gamma$. Line correspond to analytical solution of the differential equations with $\gamma=0$.}
\label{srelax}
\end{figure}

Fig. \ref{srelax} shows the numerical solution for a device with $r_\uparrow = 5$,
$r_\downarrow =1$, $\epsilon_0 = 0$ and different relaxation rates $\gamma$. The results
for $\gamma=0$ agree with analytical solution. It can be seen that relatively small spin
relaxation $\gamma=0.1$ does not alter magnetoresistance significantly even at high voltages
$\zeta$. The tendency of magnetoresistance to grow with applied voltage remains even for
the large relaxation rates $\gamma \ge 1$.

\section{Discussion}

In a fully metallic device the conductivity of the normal layer does not depend on the shift of chemical potential. The description of the device in terms of electrochemical potential does not require the knowledge of the separate contributions from electrical and chemical potentials. In our case, where the conductivity is governed by few functional levels near Fermi level, the
situation is different. Small shifts of the chemical potential change drastically the conductivity
 of the intermediate layer. Indeed, assuming the existence of the local chemical potential via the
 distribution function
\[
f_{\sigma}={1\over{\exp{(\beta(\mathcal{E}_0-\mu_{\sigma}(x)))}+1}}
\]
we may rewrite Eqs.(\ref{kinetik3}) in the standard form
\[
{J_{\sigma}\over{en}}=\sigma(\mu){d\xi_{\sigma}\over{dx}}
\]
where $\xi_{\sigma}$ is electrochemical potential and
$\sigma(\mu)=\beta e D f_{\sigma}(1-f_{\sigma})$ is
 chemical potential dependent conductivity.
Therefore our theory allows standard formulation in terms of nonlinear conductivity.

The shift of the chemical potential always occur in the spin-valve device with injection, although in devices with direct ferromagnet-normal layer contact (without contact resistance) it has slightly different form \cite{schmidt,van_Son}. Therefore the non-linear effects in magnetoresistance should be considered in all devices where the shift of the chemical potential of the order of applied voltage can significantly effect the conductivity.

Let us consider the case when the transport in the organic layer is governed by the HOMO and LUMO. The voltage $\sim 0.1$V is not sufficient to inject carriers into these states and all the carriers in the HOMO and LUMO are thermally activated. The number of such carriers is small and it allows us to neglect the space charge making our theory applicable to this case. Transport via the HOMO and LUMO corresponds to the large energy ${\cal E}_0 \sim 1eV$ and leads to the exponential suppression of the spin-valve effect with applied voltage when $eV>T$ (see also\cite{smith}). This exponential suppression was never observed in the experiments. It gives another argument (besides the ones discussed in \cite{Yu2014}) against the HOMO or LUMO transport in the organic spin-valves.

These arguments are in line with the arguments presented in Ref.\cite{Yu2014} invoking
the presence of
impurity states in the organic gap to describe the organic SV devices. The nature of these states however is far from being clear. It was reported  \cite{rybicki2012} that similar defects
may be created during sample preparation, leading to the suppression of the spin valve effect in
sufficiently thick samples. The concentration of such defects can be very low (sub-percent
level), precluding their detection by spectroscopic techniques \cite{rybicki2012}. The author of
Ref.\cite{Yu2014} considered impurity states with concentration $\sim 10^{19}cm^{-3}$ and broad distribution of energies. Such a broad distribution yields $\sigma(\mu) \approx const$ and the discussed non-linearity is absent. It was also assumed\cite{Yu2014} that the organic layer is highly conductive and its resistance is negligible relative to contact resistances. Note that the conductivity of a system of localized states with a broad energy distribution is governed by the small number of states with the energy close to the Fermi level \cite{Efr-Sh}. Therefore the actual concentration of impurities that contribute to conductivity is $\sim 10^{17}cm^{-3}$ which can hardly lead to the high conductivity of the organic layer.

There are two possible solutions of this problem. The impurity states may be arranged in some sort of conducting channels or their energy distribution may be more narrow than assumed in Ref.\cite{Yu2014}. The second option leads to the strong non-linear phenomena related to $\sigma(\mu)$ dependance. In the present paper we discussed these phenomena in the extreme case when the width of energy distribution of impurity states is less than the temperature.

\section{Conclusion}

We derived a set of kinetic equations describing non linear effects related to injection and
 transport of spin polarized carriers in organic semiconductors with hopping conductivity.
The model predicts a strongly voltage dependent magnetoresistance splitted into three distinct
regimes. The first regime (low voltages) corresponds to the well known in inorganic spintronics
conductivity mismatch limitation, underlining the correctness of the applied approach.
The second regime at intermediate voltages corresponds to fully suppressed magneto-resistance.
Interestingly, a third regime develops at higher voltages and accounts for a novel and purely
organic paradigm.
It is promoted
by strongly non-linear effects in organic semiconductor which strength is characterized by the
dimensionless parameter $eU/k_BT$. This nonlinearity, depending on device conditions, can lead to
both significant enhancement or to exponential suppression of the spin-valve effect in organic
devices. Consequently, while in inorganic devices the conductivity mismatch limitation can be
lifted off only by inserting a spin dependent tunnelling resistance,
organic devices feature an additional mechanism based on nonlinear charge transport behaviour.

We are grateful to A. Riminucci and V.I. Kozub  for stimulating discussions. This work was supported by the European Project NMP3-SL-2011-263104 "HINTS".


\begin{thebibliography}{200}
\bibitem{Dediu2009} V. Dediu, L. E. Hueso, I. Bergenti, C. Taliani, Nature Materials {\bf 8},
707716 (2009).
\bibitem{Xiong2004}  Z. H. Xiong, D. Wu, Z. V. Vardeny, J. Shi, Nature {\bf 427}, 821 (2004).
\bibitem{Xu2007}  W. Xu, G. J. Szulczewski, P. LeClair, I. Navarrete, R. Schad, G. Miao, H. Guo,
A. Gupta Appl. Phys. Lett. {\bf 90}, 072506 (2007).
\bibitem{Vinzelberg2008}  H. Vinzelberg, J. Schumann, D. Elefant, R. B. Gangineni, J. Thomas,
and B. Buchner Appl. Phys. Lett. {\bf 103}, 093720 (2008).
\bibitem{Dediu2008} V. Dediu, L. E. Hueso, I. Bergenti, A. Riminucci, F. Borgatti, P. Graziosi,
C. Newby, F. Casoli, M. P. De Jong, C. Taliani, Y. Zhan Phys. Rev. B {\bf 78}, 115203 (2008).
\bibitem{Sun2010} D. Sun, L. Yin, C. Sun, H. Guo, Z. Gai, X.-G. Zhang, T. Z. Ward, Z. Cheng,
J. Shen, Phys. Rev. Lett. {\bf 104}, 236602 (2010).
\bibitem{Barraud2010} C. Barraud, P. Seneor, R. Mattana, S. Fusil, K. Bouzehouane, C. Deranlot,
P. Graziosi, L. Hueso, I. Bergenti, V. Dediu, F. Petroff, A. Fert,
Nature Physics {\bf 6}, 615 (2010).
\bibitem{Steil2013} S. Steil, N. Grossmann, M. Laux, A. Ruffing, D. Steil, M. Wiesenmayer,
S. Mathias, O.L.A. Monti, M. Cinchetti, M. Aeschlimann, Nature Phys. {\bf 9}, 242 (2013).
\bibitem{Dediu2013} V. A. Dediu, Nature Physics {\bf 9}, 210 (2013).
\bibitem{Moodera} T. S. Santos, J. S. Lee, P. Migdal, I. C. Lekshmi, B. Satpati,
J. S. Moodera, Phys. Rev. Lett. {\bf 98}, 016601 (2007).
\bibitem{Coey} H. Tokuc, K. Oguz, F. Burke, J. M. D. Coey,
J. of Phys.: Conference Series {\bf 303}, 012097 (2011).
\bibitem{Jiang2008} J.S. Jiang, J.E. Pearson, S.D. Bader, Phys. Rev. B {\bf 77}, 035303 (2008).
\bibitem{Wang2005} F.J. Wang, Z.H. Xiong, D. Wu, J. Shi, Z.V. Vardeny,
Synthetic Metals {\bf 155}, 172 (2005).
\bibitem{Seneor_Mattana_unpub} P. Seneor, R. Mattana et al., unpublished
\bibitem{dediu_riminucci2013} V. Dediu, A. Riminucci Nature Nanotechnology {\bf 8}, 885 (2013).
\bibitem{Yu2014} Z. G. Yu, Nature Communications, {\bf 5}, 4842 (2014).
\bibitem{rybicki2012} J. Rybicki, R. Lin, F. Wang, M. Wohlgenannt, C. He, T. Sanders, and
Y. Suzuki, Phys. Rev. Lett. {\bf 109}, 076603 (2012).
\bibitem{schmidt} G. Schmidt, D. Ferrand, L.W. Molenkamp, A.T. Filip, B.J. van Wees,
Phys. Rev. B {\bf 62}, R4790 (2000).
\bibitem{van_Son} P.C. van Son, H. van Kempen, P. Wyder, Phys. Rev. Lett. {\bf 58}, 2271 (1987)
\bibitem{Fert_Jaffres} A. Fert and H. Jaffres, Phys. Rev. B {\bf 64}, 184420 (2001).

\bibitem{ValetFert} T. Valet and A. Fert, Phys. Rev. B {\bf 48}, 7099 (1993)

\bibitem{rashba} E.I. Rashba Phys. Rev. B {\bf 62}, R16267 (2000).
\bibitem{Riminucci} A. Riminucci, M. Prezioso, C. Pernechele, P. Graziosi, I. Bergenti,
R. Cecchini, M. Calbucci, M. Solzi, V. A. Dediu, Appl. Phys. Lett. {\bf 102}, 092407 (2013).
\bibitem{schmidt_hanle}  M. Grunewald, R. Gockeritz, N. Homonnay, F. Wurthner, L. W. Molenkamp,
G. Schmidt, Phys. Rev. B {\bf 88}, 085319 (2013).
\bibitem{Johnson_Silsbee} M. Johnson and R. H. Silsbee, Phys. Rev. Lett. {\bf 55}, 1790 (1985).
\bibitem{Monzon} F.G. Monzon, H.X. Tang, and M.L. Roukes, Phys. Rev. Lett. {\bf 84}, 5022 (2000).
\bibitem{Yu2013} Z. G. Yu, Phys. Rev. Lett. {\bf 111},016601 (2013).
\bibitem{smith} P. P. Ruden, D. L. Smith, Appl. Phys. Lett. {\bf 95}, 4898 (2004); A. Goswami,
M. Yunus, P. P. Ruden, and D. L. Smith, Appl. Phys. Lett. {\bf 111}, 034505 (2012).
\bibitem{deJong}
T. Lan Anh Tran, T. Quyen Le, Johnny G. M. Sanderink, Wilfred G. van der Wiel,
and Michel P. de Jong, Adv. Funct. Mat. {\bf 22}, 1180 (2012).
\bibitem{koopmans} J. J. H. M. Schoonus, P. G. E. Lumens, W. Wagemans, J. T. Kohlhepp,
P. A. Bobbert, H. J. M. Swagten, and B. Koopmans, Phys. Rev. Lett. {\bf 103}, 146601 (2009).
\bibitem{bobbert} P.A. Bobbert, W. Wagemans, F.W.A. van Oost, B. Koopmans, M. Wohlgenannt,
Phys. Rev. Lett. {\bf 102}, 156604 (2009).
\bibitem{bryksin} H. B\"ottger, V.V. Bryksin, Hopping Conduction in Solids,
Akademie-Verlag Berlin 1985.
\bibitem{YuFlatte} Z.G. Yu, M.E. Flatte, Phys. Rev. B {\bf 66}, 235302 (2002).
\bibitem{Pershin} Yu. V. Pershin, Phys. Rev. B {\bf 68}, 233309 (2003).
\bibitem{sanvito} G. Szulczewski, S. Sanvito and M. Coey, Nat. Mater., {\bf 8}, 693 (2009).

\bibitem{Efr-Sh}
B.I. Shklovskii and A.L. Efros, "Electronic Properties of Doped
Semiconductors" (Springer, Berlin, 1984).

\end{thebibliography}
\end{document}